\documentclass[prb,showpacs,preprintnumbers,amsmath,amssymb,showkeys]{revtex4}

\usepackage{epsfig}
\usepackage{graphicx}
\usepackage{dcolumn}
\usepackage{bm}
\usepackage{wasysym}
\usepackage{marvosym}
\usepackage{amsmath,amssymb}

\newcommand{\beq}{\begin{equation}} \newcommand{\eeq}{\end{equation}}
\newcommand{\ket}[1]{\mbox{$ \mid #1\, \rangle$}}
\newcommand{\bra}[1]{\mbox{$ \langle\, #1\mid$}}

\newcommand{\vecg}[1]{\mbox{${\boldsymbol #1}$}}

\newcommand{\vecb}[1]{\mbox{\bf\scriptsize #1}}

\begin{document}

\title{Evolution of squeezed states under the Fock-Darwin Hamiltonian}
\author{Jaime E. Santos and Nuno M. Peres}
\affiliation{Centro de F\'isica and Departamento de F\'isica, Universidade do Minho, P-4710-057 Braga, Portugal}
\email{jaime.santos@fisica.uminho.pt}
\author{Jo\~ao Lopes dos Santos}
\affiliation{CFP and Departamento de F\'isica, Faculdade de Ci\^encias, Universidade do Porto, 4169-007 Porto, Portugal}

\date{\today}

\begin{abstract}
We develop a complete analytical description of the time evolution of
squeezed states of a charged particle under the Fock-Darwin 
Hamiltonian and a time-dependent electric field. This result generalises 
a relation obtained by Infeld and Pleba\'nski for states of the 
one-dimensional harmonic oscillator.
We relate the evolution of a state-vector subjected
to squeezing to that of state which is not subjected to squeezing
and for which the time-evolution under the simple harmonic oscillator dynamics
is known (e.g. an eigenstate of the
Hamiltonian). A corresponding relation is also established for the
Wigner functions of the states, in view of
their utility in the analysis of cold-ion experiments. In an
appendix, we compute the response functions of the FD Hamiltonian to
an external electric field, using the same techniques as in the main
text.
\end{abstract}

\maketitle

\section{Introduction}
\label{secA}

The study of the time-evolution of a non-relativistic charged particle 
in homogeneous magnetic and electric fields has a long history in physics.
In a quantum context, the treatment of the problem goes back to Darwin
\cite{Darwin27}, who considered the evolution of a Gaussian wave-packet
in a magnetic field, and Fock \cite{Fock28} who obtained the eigenenergies
and eigenstates of a charged particle in an isotropic harmonic potential, 
subjected to a magnetic field normal to the plane of motion. 

If one takes a particle of charge $-e$ and mass $m$, moving in the $xy$
plane in an harmonic potencial of frequency $\omega_0$ 
and subjected to a magnetic field $\vecg{B}=B\vecg{e}_z$, the Hamiltonian describing the system 
is given, in the symmetric gauge where
$\vecg{A}=\frac{1}{2}\,\vecg{B}\times\vecg{r}$, by
\begin{equation}
\hat{H}_0=\frac{1}{2m}\,\left[\left(\hat{p}_x-\frac{eB}{2}\hat{y}\right)^2+
\left(\hat{p}_y+\frac{eB}{2}\hat{x}\right)^2\right]+\frac{1}{2}\,m\,\omega_0^2\,(\hat{x}^2+
\hat{y}^2)\,,
\label{eq1}
\end{equation}
where the operators $\hat{x},\hat{p}_x,\hat{y},\hat{p}_y$ obey the canonical commutation relations.

This simple problem has applications in the context of
the Quantum Hall effect \cite{Hajdu94}, where disorder and the Coulomb
interaction also play a crucial role. Another field for which the
study of this Hamiltonian has proved fruitful is that of quantum
dots, where the simple Hamiltonian given by (\ref{eq1}) seems to give a
good account of the $I-V_g$ curves obtained when a gate voltage is
applied to the quantum dot \cite{Kouwenhoven01}, with corrections
due to the assymetry of the confining potential and to the Coulomb
interactions also playing a role. For some types of quantum dots,
such as InAs/GaAs quantum dots, the agreement between the theoretical and
experimental results seems to hold for magnetic fields up to 15 T 
\cite{Babinski06}.

The study of the evolution of Gaussian wave packets also
goes back to the first days of quantum mechanics. This study was first
undertaken by Schr\"odinger \cite{Schrodinger26}, Kennard
\cite{Kennard27} and also by Darwin \cite{Darwin27} in the context of
the harmonic oscillator, of a free particle and of a particle in
constant electric and magnetic fields. This problem
continues to attract attention to the present day in many contexts,
see the review by Dodonov \cite{Dodonov02}.

Schr\"odinger considered the time-evolution of a minimal uncertainty
state, i.e. a coherent state of the harmonic
oscillator, in the terminology of Glauber \cite{Glauber63}. These
states have a wide range of applications in quantum optics (see e.g.
\cite{Leonhardt97}), where they act as the quasi-classical states of the
EM field, and in quantum field theory, where they are the basis of the
phase-space path integral \cite{Brown94}. Such states and
their derivatives have become important in quantum information
processing in recent years, in the context of the manipulation of
cold atoms in traps. It has become possible to reconstruct the
Wigner function of a coherent state of the center-of-mass of an 
harmonically bound ion \cite{Leibfried96}.

Kennard has on the other hand considered the evolution of a more
general wavepacket of the harmonic oscillator, what is now known
as a squeezed state. Important 
early contributions are those of Husimi \cite{Husimi53} and
Infeld and Pleba\'nski \cite{Infeld55,Plebanski56}. These later authors
introduced the so-called squeezing operator and established a
relation between the evolution of initial states
for which the time-evolution is known ('unsqueezed states') 
and states which are derived from such initial states by the application of the
squeezing operator ('squeezed states'). The generalisation of this relation to
the FD Hamiltonian in an homogeneous time-dependent electric 
field is the main result of the present paper. Stoler \cite{Stoler70,Stoler71} 
proved that squeezed-coherent states are unitarily equivalent to 
coherent states, thus being minimal uncertainty states, but that 
their minimal character is not
preserved by time-evolution, although the uncertainty periodically
assumes the minimum value compatible with Heisenberg's uncertainty 
relation. He also showed that the squeezing operator as currently
written in terms of quadratures operators is exactly of the form
given by Infeld and Pleba\'nski. Squeezed states play a proeminent role
in quantum optics, see again \cite{Leonhardt97}. Recently, 
they have also become important in quantum information processing
both in a quantum optics context and also through the manipulation of
cold atoms \cite{Heinzen90,Cirac93,Zeng95,Meekhof96,Ban99,Braunstein00,
Braunstein05,Vahlbruch07,Polzik08,Adesso08}. See in particular reference 
\cite{Kimble08} for an up-to-date report of the state
of quantum information processing using cold atoms and photons.

Besides the paper of Darwin already referred, the time-evolution of
states of a non-relativistic particle in homogeneous electric and 
magnetic fields was also considered by Malkin, Man'ko, Trifonov
and Dodonov \cite{Malkin69,Malkin69a,Malkin70,Malkin70a,Dodonov72}, 
who considered the dynamics of a particle in an homogeneous
electromagnetic field in terms of coherent states, obtaining an 
explicit representation of the Green's function, and studied the 
invariants of the system; by 
Lewis and Riesenfeld \cite{Lewis69}, who also considered the
invariants of such a system; by Kim and Weiner \cite{Kim73},
who considered the evolution of Gaussian wave-packets in 
a magnetic field, subjected to an isotropic harmonic potential
(i.e. the FD Hamiltonian), but also to sadle-point potentials, which
are relevant for tunnelling problems. 

The structure of this paper is as follows: in the next section, we
review the notion of squeezing operator and generalise the relation
of Infeld and Pleba\'nski to states evolving under the Fock-Darwin
Hamiltonian. In section \ref{secC}, with a view to applications in
the manipulation of cold atoms, we use the relation obtained to
establish a relation between the Wigner function of different
states and apply it to the special case of squeezed-coherent states. 
In section \ref{secD}, we present our conclusions.
In appendix \ref{appendixA}, we present a derivation of the
finite frequency permitivity and conductivity of the FD Hamiltonian
that uses the same operator methods that are used in the main text
but which lies somewhat outside of the scope of the main
text. Finally, in appendix \ref{appendixB}, we derive the
original Infeld-Pleba\'nski relation through elementary means.  

\section{Evolution of squeezed states under the FD Hamiltonian}
\label{secB}

We consider the time-evolution of the state $\ket{\overline{\psi}_t}$,
that obeys the time-dependent Schr\"odinger equation
$i\hbar\,\partial_t\,\ket{\overline{\psi}_t}=\hat{H}(t)\,\ket{\overline{\psi}_t}$,
where $\hat{H}(t)=\hat{H}_0+\hat{H}_1(t)$, with $\hat{H}_0$ being given
by equation (\ref{eq1}) and where the interaction 
Hamiltonean $\hat{H}_1(t)$ of the charge with the external electric
field is given, in the Schr\"odinger picture, by $\hat{H}_1(t)=e(E_x(t)\hat{x}+E_y(t)\hat{y})$.

If one now expands the squares and groups the different terms of (\ref{eq1}), one obtains
\begin{equation}
\hat{H}_0=\frac{1}{2m}\,\left(\hat{p}_x^2+
\hat{p}_y^2\right)+\frac{1}{2}\,m\,\omega_R^2\,(\hat{x}^2+
\hat{y}^2)+\frac{\omega_L}{2}\,\hat{L}_z\,,
\label{eq2}
\end{equation}
where $\omega_L=\frac{eB}{m}$ is the giration frequency and 
$\omega_R^2=\omega_0^2+\frac{\omega_L^2}{4}$,
with $\hat{L}_z=\hat{x}\hat{p}_y-\hat{y}\hat{p}_x$ being the angular
momentum component along the $z$ axis. 
One should note that one can write
$\hat{H}_0=\hat{h}_0+\frac{\omega_L}{2}\,\hat{L}_z$, where $\hat{h}_0$
is the Hamiltonean of the isotropic harmonic oscillator with frequency
$\omega_R$, and also that $[\hat{h}_0,\hat{L}_z]=0$.

Given the state vector $\ket{\overline{\psi}_t}$, one can define
the corresponding state vector $\ket{\phi_t}=e^{\frac{i}{\hbar}\,\hat{H}_0
  t}\,\ket{\overline{\psi}_t}$, in the interaction representation, such that
the two vectors coincide at $t=0$. This state vector evolves acording to 
the Hamiltonian $\hat{H}_1^{int}(t)=
e^{\frac{i}{\hbar}\,H_0t}\,\hat{H}_1(t)\,e^{-\frac{i}{\hbar}\,H_0t}$, which given that 
$\hat{h}_0$ and $\hat{L}_z$ in $\hat{H}_0$ commute, one can also write as
\begin{equation}
\hat{H}_1^{int}(t)=e^{\frac{i}{\hbar}\,\hat{h}_0t}\,e^{\frac{i\omega_Lt}{2\hbar}\,\hat{L}_z}
\,\hat{H}_1(t)\,e^{-\frac{i\omega_Lt}{2\hbar}\,\hat{L}_z}\,e^{-\frac{i}{\hbar}\,\hat{h}_0t}\,.
\label{eq2.1}
\end{equation}
If one now applies the time-dependent rotation, encoded by
$\hat{L}_z$ to $\hat{H}_1(t)$, followed by the dynamics of the isotropic 
harmonic oscillator, encoded in $\hat{h}_0$, one obtains for $\hat{H}_1^{int}(t)$:
\begin{eqnarray}
\hat{H}_1^{int}(t)&=&e\left\{E_{x}'(t)\left[\,\hat{x}\,\cos(\omega_R
  t)+\hat{p}_x\,\sin(\omega_R t)\,\right]+\,E_{y}'(t)\left[\,\hat{y}\,\cos(\omega_R
  t)+\hat{p}_y\,\sin(\omega_R t)\,\right]\right\},
\label{eq3}
\end{eqnarray}
where 
\begin{eqnarray}
E_{x}'(t)&=&E_x(t)\cos\left(\frac{\omega_L}{2}\,t\right)
+E_y(t)\sin\left(\frac{\omega_L}{2}\,t\right)\,,
\label{eq3.1a}\\
E_{y}'(t)&=&-E_x(t)\sin\left(\frac{\omega_L}{2}\,t\right)+E_y(t)\cos\left(\frac{\omega_L}{2}\,t\right)\,,
\label{eq3.1b}
\end{eqnarray}
are the components of the electric field in the rotated frame.

The wave equation for $\ket{\phi_t}$ can be formally integrated in 
terms of time-ordered products of the integral of
$\hat{H}_1^{int}(t)$, i.e. $\ket{\phi_t}=T\,\exp\left(-\frac{i}{\hbar}
\,\int_{0}^{t}\,du\,\hat{H}_1^{int}(u)\right)\,
\ket{\overline{\psi}_{0}}$. Now, since the commutator of the 
operators $\hat{H}_1^{int}(u)$ at different times is a c-number,
one can write the time ordered operator above as 
\begin{eqnarray}
T\,\exp\left(-\frac{i}{\hbar}\,\int_{0}^{t}\,du\,\hat{H}_1^{int}(u)\right)&=&
\exp\left(-\frac{i}{\hbar}\,\int_{0}^{t}\,du\,\hat{H}_1^{int}(u)\right)\nonumber\\
&&\mbox{}\times\,\exp\left(-\frac{1}{2\hbar^2}\,\int_{0}^t\,du\,\int_{0}^{u}\,dv\,
[\hat{H}_1^{int}(u),\hat{H}_1^{int}(v)]\right)\,,
\label{eq3_1}
\end{eqnarray}
where the second term on the rhs is a phase factor.

Collecting the several terms, one obtains for the evolution of $\ket{\overline{\psi}_t}$ 
\begin{eqnarray}
\ket{\overline{\psi}_t}&=&\exp\left(\frac{ie^2}{2\hbar
    m\omega_R}\,\int_{0}^t\,du\,\int_{0}^{u}\,dv\,\sin[\,\omega_R(u-v)\,]\,
\,[E_x'(u)E_x'(v)+E_y'(u)E_y'(v)]\right)\nonumber\\
&&\mbox{}\times
\exp\left(-\frac{i\omega_Lt}{2\hbar}\,\hat{L}_z\right)\,
\exp\left(-\frac{i}{\hbar}\,\hat{h}_0\,t\right)\,\exp\left(-\frac{i}{\hbar}
\,\int_{0}^{t}\,du\,\hat{H}_1^{int}(u)\right)\,
\ket{\overline{\psi}_0}\,.
\label{eq4}
\end{eqnarray}
One now assumes, following Infeld and Pleba\'nski \cite{Plebanski56}, that the initial
state $\ket{\overline{\psi}_0}$ is related to a certain initial state
$\ket{\psi_0}$, for which the time-evolution under the isotropic harmonic
oscillator Hamiltonian $\hat{h}_0$ is known, by
\begin{equation}
\ket{\overline{\psi}_0}=\exp\left(\frac{i}{\hbar}(\,
P^0_x\hat{x}+P^0_y\hat{y}-X_0\hat{p}_x-Y_0\hat{p}_y\,)\right)\,
\exp\left(\frac{i}{2\hbar}\,r\,(\hat{\vecg{p}}\cdot\hat{\vecg{r}}+
\hat{\vecg{r}}\cdot\hat{\vecg{p}})\right)\,\ket{\psi_0}\,,
\label{eq5}
\end{equation}
where the first operator is a translation operator in phase space, with
$X_0,Y_0,P_x^0,P_y^0$ being arbitrary real constants, and the second
operator is the squeezing operator, with $r$ being a real constant 
that indicates the degree of squeezing.

Substituting equation (\ref{eq5}) in (\ref{eq4}) one can combine 
the operators
$e^{-\frac{i}{\hbar}\,\int_{0}^{t}\,du\,\hat{H}_1^{int}(u)}$ and
$e^{\frac{i}{\hbar}(\,P^0_x\hat{x}+P^0_y\hat{y}-X_0\hat{p}_x-Y_0\hat{p}_y\,)}$,
since the commutator of their exponents is a c-number. This operation
merely generates a phase factor, coming from the commutator. One can
then commute the resulting operator to the left-hand side, through
$e^{-\frac{i\omega_Lt}{2\hbar}\,\hat{L}_z}\,e^{-\frac{i}{\hbar}\,\hat{h}_0t}$,
using the time-evolution of $\hat{x},\hat{p}_x,\hat{y}, \hat{p}_y$
under $\hat{h}_0$ and under the time-dependent rotation around the $z$
axis. Combining the resulting phase factors and operators, we obtain
\begin{eqnarray}
\ket{\overline{\psi}_t}&=&\exp\left(-\frac{ie}{2\hbar}\,\int_{0}^t\,du\,\,
[E_x(u)\,x_c(u)+E_y(u)\,y_c(u)]\right)\,
\exp\left(\frac{i}{\hbar}(\,p^c_x(t)\hat{x}+p^c_y(t)\hat{y}
-x_c(t)\hat{p}_x-y_c(t)\hat{p}_y\,)\right)\nonumber\\
&&\mbox{}\times
\exp\left(-\frac{i\omega_Lt}{2\hbar}\,\hat{L}_z\right)\,
\exp\left(-\frac{i}{\hbar}\,\hat{h}_0\,t\right)\,
\exp\left(\frac{i}{2\hbar}\,r\,(\hat{\vecg{p}}\cdot\hat{\vecg{r}}+
\hat{\vecg{r}}\cdot\hat{\vecg{p}})\right)\,\ket{\psi_0}\,.
\label{eq6}
\end{eqnarray}
where $x_c(t)$, $y_c(t)$, $p_x^c(t)$ and $p_y^c(t)$ are the classical
solutions of the equations of motion for the Fock-Darwin problem with
initial positions $X_0$ and $Y_0$ and initial momenta $P_x^0$
and $P_y^0$. These solutions are given by
\begin{eqnarray}
x_c(t)&=&\left[\,X_0\cos\left(\frac{\omega_L}{2}t\right)
-Y_0\sin\left(\frac{\omega_L}{2}t\right)\,\right]\,\cos(\omega_Rt)\nonumber\\
&&\mbox{}+\frac{1}{m\omega_R}\,\left[\,P_x^0
\cos\left(\frac{\omega_L}{2}t\right)-P_y^0
\sin\left(\frac{\omega_L}{2}t\right)\,\right]\,\sin(\omega_Rt)\nonumber\\
&&\mbox{}-\frac{e}{2m\omega_R}\,\int_0^t\,du\,
\left\{E_x(u)\,\left[\sin(\omega_+(t-u))+\sin(\omega_-(t-u))\right]\right.
\nonumber\\
&&\mbox{}\left.+E_y(u)\,\left[\cos(\omega_+(t-u))-\cos(\omega_-(t-u))\right]\right\}\,,
\label{eq7}
\end{eqnarray}
\begin{eqnarray}
y_c(t)&=&\left[\,X_0\sin\left(\frac{\omega_L}{2}t\right)+Y_0\cos\left(\frac{\omega_L}{2}t\right)\,\right]\,\cos(\omega_Rt)\nonumber\\
&&\mbox{}+\frac{1}{m\omega_R}\,\left[\,P_x^0
\sin\left(\frac{\omega_L}{2}t\right)+P_y^0
\cos\left(\frac{\omega_L}{2}t\right)\,\right]\,\sin(\omega_Rt)\nonumber\\
&&\mbox{}-\frac{e}{2m\omega_R}\,\int_0^t\,du\,\left\{-E_x(u)\,\left[\cos(\omega_+(t-u))-\cos(\omega_-(t-u))\right]\right.
\nonumber\\
&&\mbox{}\left.+E_y(u)\,\left[\sin(\omega_+(t-u))+\sin(\omega_-(t-u))\right]\right\}\,,
\label{eq8}
\end{eqnarray}
and 
\begin{eqnarray}
p_x^c(t)&=&\left[\,P_x^0\cos\left(\frac{\omega_L}{2}t\right)-P_y^0\sin\left(\frac{\omega_L}{2}t\right)\,\right]\,\cos(\omega_Rt)\nonumber\\
&&\mbox{}-m\omega_R\,\left[\,X_0
\cos\left(\frac{\omega_L}{2}t\right)-Y_0
\sin\left(\frac{\omega_L}{2}t\right)\,\right]\,\sin(\omega_Rt)\nonumber\\
&&\mbox{}-\frac{e}{2}\,\int_0^t\,du\,\left\{E_x(u)\,\left[\cos(\omega_+(t-u))+\cos(\omega_-(t-u))\right]\right.
\nonumber\\
&&\mbox{}\left.-E_y(u)\,\left[\sin(\omega_+(t-u))-\sin(\omega_-(t-u))\right]\right\}\,,
\label{eq9}
\end{eqnarray}
\begin{eqnarray}
p_y^c(t)&=&\left[\,P_x^0\sin\left(\frac{\omega_L}{2}t\right)+P_y^0\cos\left(\frac{\omega_L}{2}t\right)\,\right]\,\cos(\omega_Rt)\nonumber\\
&&\mbox{}-m\omega_R\,\left[\,X_0
\sin\left(\frac{\omega_L}{2}t\right)+Y_0
\cos\left(\frac{\omega_L}{2}t\right)\,\right]\,\sin(\omega_Rt)\nonumber\\
&&\mbox{}-\frac{e}{2}\,\int_0^t\,du\,\left\{E_x(u)\,\left[\sin(\omega_+(t-u))-\sin(\omega_-(t-u))\right]\right.
\nonumber\\
&&\mbox{}\left.+E_y(u)\,\left[\cos(\omega_+(t-u))+\cos(\omega_-(t-u))\right]\right\}\,.
\label{eq10}
\end{eqnarray}

One can read the classical dielectric permitivity of the system from
equations (\ref{eq7},\ref{eq8}) and 
one obtains the same results as in appendix \ref{appendixA}, which
shows that the quantum and classical results are 
identical, as one would expect for a linear system. The classical
velocities $v_x^c(t), v_y^c(t)$ can be easily computed by derivation of
(\ref{eq7},\ref{eq8}) with respect to time or, using the relations
$v_x^c(t)=\frac{1}{m}\left(p_x^c(t)-\frac{eB}{2}\,y_c(t)\right)$ and 
$v_y^c(t)=\frac{1}{m}\left(p_y^c(t)+\frac{eB}{2}\,x_c(t)\right)$, from equations
(\ref{eq7}) to (\ref{eq10}). One can then read the classical conductivity
of the system from the resulting expression and this result again
coincides with that of appendix \ref{appendixA}, 
i.e. the classical and quantum results are the same. Finally, one can
also show, after derivation of 
the velocity expressions with respect to time, that the solution given 
by (\ref{eq7}) and (\ref{eq8}) obeys the classical equations of
motion in two dimensions $\ddot{\vecg{r}}=-
\frac{e}{m}(\vecg{E}(t)+\,\dot{\vecg{r}}\times \vecg{B})-\omega_0^2\,\vecg{r}$,
with $\vecg{B}=B\,\vecg{e}_z$.

The left most operator in equation (\ref{eq6}) is again a translation
operator in phase phase. If one considers the wave-function in the 
coordinate representation,
$\overline{\psi}(x,y,t)=\langle\,x,y\,\ket{\overline{\psi}_t}$,
and one applies this translation operator to $\bra{x,y}$, followed by
the rotation operator $e^{-\frac{i\omega_Lt}{2\hbar}\,\hat{L}_z}$, one obtains the following result
\begin{eqnarray}
\overline{\psi}(x,y,t)&=&\exp\left(-\frac{i}{2\hbar}\left(\,p_x^c(t)\,x_c(t)+p_y^c\,y_c(t)
+e\int_{0}^t\,du\,[E_x(u)\,x_c(u)+E_y(u)\,y_c(u)]\,\right)\right)\nonumber\\
&&\mbox{}\times\exp\left(\frac{i}{\hbar}[p^c_x(t)\,x+p^c_y(t)\,y]\right)\,
\bra{{\cal R}^{-1}_t\cdot\,(\,\vecg{r}-\vecg{r}_c(t)\,)} \,\,e^{-\frac{i}{\hbar}\,\hat{h}_0\,t}\,
e^{\frac{i}{2\hbar}\,r\,(\hat{\vecb{p}}\cdot\hat{\vecb{r}}+
\hat{\vecb{r}}\cdot\hat{\vecb{p}})}\,\ket{\psi_0}\,,
\label{eq11}
\end{eqnarray}
with $\vecg{r}_c(t)$ being the classical solutions of the 
equations of motion (\ref{eq7}) and (\ref{eq8}) and ${\cal R}_t$ being
the rotation matrix in two dimensions by an angle of $\omega_Lt/2$.

For the wave-function in the momentum representation $\overline{\psi}(p_x,p_y,t)=\langle\,p_x,p_y\,\ket{\overline{\psi}_t}$, one
obtains, applying the translation operator to $\bra{p_x,p_y}$,
followed again by the rotation operator,
\begin{eqnarray}
\overline{\psi}(p_x,p_y,t)&=&
\exp\left(\frac{i}{2\hbar}\left(\,p_x^c(t)\,x_c(t)+p_y^c\,y_c(t)
-e\int_{0}^t\,du\,[E_x(u)\,x_c(u)+E_y(u)\,y_c(u)]\,\right)\right)\nonumber\\
&&\mbox{}\times\exp\left(-\frac{i}{\hbar}[p_x\,x_c(t)+p_y\,y_c(t)]\right)\,
\bra{{\cal R}^{-1}_t\cdot\,(\,\vecg{p}-\vecg{p}^c(t)\,)} \,\,e^{-\frac{i}{\hbar}\,\hat{h}_0t}\,e^{\frac{i}{2\hbar}\,r\,(\hat{\vecb{p}}\cdot\hat{\vecb{r}}+
\hat{\vecb{r}}\cdot\hat{\vecb{p}})}\,\ket{\psi_0}\,,
\label{eq12}
\end{eqnarray}
with $\vecg{p}^c(t)$ being the classical solutions of the 
equations of motion (\ref{eq9}) and (\ref{eq10}).

One can now apply the relation of Infeld and Pleba\'nski
\cite{Plebanski56}, relating the evolution of $\bra{x,y}
\,\,e^{-\frac{i}{\hbar}\,\hat{h}_0t}
\,e^{\frac{i}{2\hbar}\,r\,(\hat{\vecb{p}}\cdot\hat{\vecb{r}}+
\hat{\vecb{r}}\cdot\hat{\vecb{p}})}\,\ket{\psi_0}$ to that of 
$\bra{x,y} \,\,e^{-\frac{i}{\hbar}\,\hat{h}_0t}\,
\ket{\psi_0}$ and the evolution of
$\bra{p_x,p_y} \,\,e^{-\frac{i}{\hbar}\,\hat{h}_0t}
\,e^{\frac{i}{2\hbar}\,r\,(\hat{\vecb{p}}\cdot\hat{\vecb{r}}+
\hat{\vecb{r}}\cdot\hat{\vecb{p}})}\,\ket{\psi_0}$ to that of 
$\bra{p_x,p_y} \,\,e^{-\frac{i}{\hbar}\,\hat{h}_0t}\,\ket{\psi_0}$, to equations
(\ref{eq11}) and (\ref{eq12}), respectively. This relation is derived
by elementary means in appendix \ref{appendixB}.

One obtains for (\ref{eq11}) the result
\begin{eqnarray}
\overline{\psi}(x,y,t)&=&\mu_r^{-1}(t)\,\exp\left(-\frac{i}{2\hbar}\,\left(p_x^c(t)
\,x_c(t)+p_y^c\,y_c(t)\,+\,e\int_{0}^t\,du\,\,[E_x(u)\,x_c(u)+E_y(u)\,y_c(u)]\right)\right)
\nonumber\\
&&\mbox{}\times
\exp\left(\frac{i}{\hbar}(p^c_x(t)\,x+p^c_y(t)\,y)\right)\,
\exp\left(\,\frac{i\,m\,\omega_R\,\sinh(2r)\sin(2\omega_Rt)}{2\,\hbar\,\mu_r^{2}(t)}\,
[\,(x-x_c(t))^2+(y-y_c(t))^2\,]\,\right)\nonumber\\
&&\mbox{}\times
\psi\left(\frac{{\cal R}^{-1}_t\cdot\,(\,\vecg{r}-\vecg{r}_c(t)\,)}{\mu_r(t)},\tau_r\right)\,,
\label{eq13}
\end{eqnarray}
where $\mu_r(t)=\sqrt{e^{-2r}\cos^2(\omega_Rt)+e^{2r}\sin^2(\omega_Rt)}$ and
$\tau_r=\frac{1}{\omega_R}\,\arctan(e^{2r}\tan(\omega_Rt))$.

For equation (\ref{eq12}), one obtains
\begin{eqnarray}
\overline{\psi}(p_x,p_y,t)&=&\mu_{-r}^{-1}(t)\,
\exp\left(\frac{i}{2\hbar}\,\left(p_x^c(t)\,x_c(t)+p_y^c\,y_c(t)\,
-\,e\int_{0}^t\,du\,\,[E_x(u)\,x_c(u)+E_y(u)\,y_c(u)]\right)\right)
\nonumber\\
&&\mbox{}\times
\exp\left(-\frac{i}{\hbar}[p_x\,x_c(t)+p_y\,y_c(t)]\right)
 \exp\left(-\frac{i\,\sinh(2r)\sin(2\omega_Rt)}{2\,\hbar \,m\,\omega_R\,\mu_{-r}^{2}(t)}\,
[\,(p_x-p_x^c(t))^2+(p_y-p_y^c(t))^2\,]\right)\nonumber\\
&&\mbox{}\times\psi\left(\frac{{\cal R}^{-1}_t\cdot\,(\,
\vecg{p}-\vecg{p}^c(t)\,)}{\mu_{-r}(t)},\tau_{-r}\right)\,,
\label{eq14}
\end{eqnarray}
where $\mu_{-r}(t)=\sqrt{e^{2r}\cos^2(\omega_Rt)+e^{-2r}\sin^2(\omega_Rt)}$ and
$\tau_{-r}=\frac{1}{\omega_R}\,\arctan(e^{-2r}\tan(\omega_Rt))$. The expressions
(\ref{eq13}) and (\ref{eq14}) constitute the main result of this paper and generalise
those of Infeld and Pleba\'nski as presented in \cite{Plebanski56}
to states evolving under the Fock-Darwin Hamiltonian
subjected to a time-dependent electric field in the plane of the system.

\section{The Wigner function for a squeezed state}
\label{secC} 

One can also establish a relation between the Wigner function
of a state subjected to squeezing in the presence of electric
and magnetic fields and the Wigner function of an 'unsqueezed state'
evolving under the isotropic harmonic oscillator dynamics. 

The Wigner function is defined as \cite{Wigner32}
\begin{equation}
D(\vecg{r},\vecg{p},t)=\frac{1}{(2\pi\hbar)^2}\,\int\,d^2v 
\,e^{-\frac{i}{\hbar}\vecb{p}\cdot\vecb{v}}\,
\overline{\psi}^*(\vecg{r}-\vecg{v}/2,t)\,
\overline{\psi}(\vecg{r}+\vecg{v}/2,t)\,.
\label{eq15}
\end{equation}
The Wigner function gives, after integration with respect to the
momentum $\vecg{p}$ or to the coordinate $\vecg{r}$,
the coordinate or momentum distribution function of the system and can be considered
a quantum generalisation of the Boltzmann distribution
function. However, the Wigner function is not a {\it bonna-fide}
distribution since it can assume negative values.

Substituting the result of equation (\ref{eq13}) in (\ref{eq15}),
one obtains, after a substitution of variable on $\vecg{v}$
\begin{equation}
D(\vecg{r},\vecg{p},t)=D_0\left(\frac{{\cal R}^{-1}_t\cdot[\vecg{r}-\vecg{r}_c(t)]}{\mu_r(t)},
{\cal R}^{-1}_t\cdot\vecg{u}(\vecg{r},\vecg{p},t),\tau_r\right)\,,
\label{eq16}
\end{equation}
where $\vecg{u}(\vecg{r},\vecg{p},t)=\mu_r(t)[\vecg{p}-\vecg{p}^c(t)]
-m\omega_R\sinh(2r)\sin(2\omega_Rt)
\frac{\vecg{r}-\vecg{r}_c(t)}{\mu_r(t)}$
and where $D_0(\vecg{r},\vecg{p},t)$ is the Wigner function of the 'unsqueezed'
initial state $\psi_0(x,y)$ evolving under the isotropic harmonic oscillator
dynamics in the absence of magnetic or electric fields.

If the 'unsqueezed' initial state $\ket{\psi_0}$ in equation
(\ref{eq5}) is the vacuum of the isotropic harmonic oscillator, then
the state $\ket{\overline{\psi}_0}$ is a squeezed-coherent state,
evolving under the Fock-Darwin Hamiltonian. In the context of 
cold ion experiments, one can produce squeezed states of the ion
center of mass, either by queenching of the trap frequency or
parametric  amplification \cite{Heinzen90}, as well as
multichromatic excitation of the ion \cite{Cirac93,Zeng95,Meekhof96}.

In this case, $D_0(\vecg{r},\vecg{p},t)=\frac{1}{\pi^2\hbar^2}\,e^{-\frac{m\omega_R}{\hbar}\vecb{r}^2
-\frac{\vecb{p}^2}{\hbar m\omega_R}}$, and we obtain for the Wigner
function of the squeezed-coherent state the result
\begin{equation}
D(\vecg{r},\vecg{p},t)=\frac{1}{\pi^2\hbar^2}\,e^{-\frac{m\omega_R \mu_{-r}^2(t)}{\hbar}\, (\vecb{r}-\vecb{r}_c(t))^2
-\frac{\mu_{r}^2(t)}{\hbar m\omega_R}\,(\vecb{p}-\vecb{p}^c(t))^2+\frac{2}{\hbar}\sinh(2r)\,\sin(2\omega_Rt)\,
(\vecb{p}-\vecb{p}^c(t))\cdot (\vecb{r}-\vecb{r}_c(t))}\,, 
\label{eq17}
\end{equation} 
which represents an assymetric Gaussian, whose shape is preserved by
the dynamics of the system, rotating in phase-space, with its center
determined by the solutions of the classical equations of motion (\ref{eq7}-\ref{eq10}).
Note that for $r=0$, i.e. for a coherent state, the Wigner function
decouples into a product of functions that depend only on the
coordinates or the momenta. Also, note that for the result
(\ref{eq17}) is strictly positive, but this is not the case in
general, e.g. if we had taken $\ket{\psi_0}$ to be an
excited-state of the harmonic oscillator. 

A group of still pictures of the $x,p_x$ section of the Wigner function (\ref{eq17})
is shown in the figures, for a system subjected to a right-handed circularly
polarised wave, incident along the $zz$ axis and aligned with the $xx$ axis at $t=0$.
The intensity of the field is $E_0=100$ Vm$^{-1}$, with frequency
equal to $\Omega=1.42\times 10^9$ Hz. The harmonic frequency of the 
trap is $\omega_0=7.04\times 10^7$ Hz and the mass of the 
particle $m=1.50\times 10^{-26}$ Kg, is that of a $^9Be^+$ ion, with the 
applied magnetic field being $B=6.60$ T, which gives a giration 
frequency $\omega_L=7.06\times 10^7$ Hz and $\omega_R=7.88\times
10^7$ Hz. Finally, the squeezing parameter $r=0.35$.

The integration of expression (\ref{eq17}) with respect to $\vecg{p}$ or
$\vecg{r}$ yields the coordinate or momentum
distribution. The wave-packets in coordinate or position space are
centered around the classical solutions of the equations of motion,
the uncertainty in e.g. $x$, $p_x$ being given by
$\langle \Delta x^2\rangle_t= \frac{\hbar}{2m\omega_R}\,
(\cos^2(\omega_Rt)e^{-2r}+\sin^2(\omega_Rt)e^{2r})$, $\langle \Delta
p_x^2\rangle_t = 
\frac{\hbar
  m\omega_R}{2}\,(\cos^2(\omega_Rt)e^{2r}+\sin^2(\omega_Rt)e^{-2r})$,
i.e. the uncertainties oscillate with period $2\pi/\omega_R$. 
Their product is given by
\begin{equation}
\langle \Delta x^2\rangle_t\,\langle \Delta p_x^2\rangle_t=\frac{\hbar^2}{4}\,(1+\sinh^2(2r)\,\sin^2(2\omega_Rt))\geq \frac{\hbar^2}{4}\,,
\label{eq18}
\end{equation}
in agreement with Heisenberg's uncertainty relation. Note that the 
uncertainty in $x$ and $p_x$ oscillate
in opposition, i.e. one increases while the other is decreasing. 
Also note that, unlike a coherent state, the
squeezed-coherent state is not a minimum uncertainty state 
for these two canonical variables, except when 
$t=\frac{n\pi}{2\omega_R}$ \cite{Stoler71}.
\section{Conclusions}
\label{secD}
In this paper, we have considered the time-evolution of general
squeezed states evolving under the Fock-Darwin Hamiltonian in an
homogeneous time-dependent electric field. We have generalised a 
relation of Infeld and Pleba\'nski, between the time-evolution of
states of the harmonic oscillator subjected
to squeezing and states not subjected to squeezing and for which the
time-evolution is known, to states evolving under the FD Hamiltonian 
in two dimensions. A corresponding relation was also established for 
the Wigner functions of the states. Finally, in an appendix, we
computed the response functions of the FD Hamiltonian to
an external electric field, using the same techniques as in the main
text.

{\bf Acknowledgements:}
We acknowledge helpful dis\-cus\-sions 
with A. Ferreira. The authors were supported by  FCT
under Grant No. PTDC/FIS/64404/2006.
\appendix
\section{Calculation of the dielectric permitivity and 
optical conductivity}
\label{appendixA}
We will first indicate how to diagonalise the 
Hamiltonian (\ref{eq1})
in the operator representation. The diagonalisation of this Hamiltonian
and the determination of its eigenfunctions was first 
obtained by Fock \cite{Fock28}. 

Introducing the annihilation and creation operators 
$\hat{a}_x,\hat{a}_y, \hat{a}_x^\dagger,
\hat{a}_y^\dagger$ through $\hat{x}=\left(\hbar/2m\omega_R
\right)^{1/2}\,(\hat{a}_x+\hat{a}_x^\dagger)$, 
$\hat{p}_x=i\left(\hbar m\omega_R/2\right)^{1/2}
\,(\hat{a}_x-\hat{a}_x^\dagger)$,
 $\hat{y}=\left(\hbar/2m\omega_R\right)^{1/2}
\,(\hat{a}_y+\hat{a}_y^\dagger)$, 
$\hat{p}_y=i\left(\hbar m\omega_R/2\right)^{1/2}
\,(\hat{a}_y-\hat{a}_y^\dagger)$, one can write such Hamiltonian
in the form
\begin{equation}
\hat{H}_0=\hbar\,\omega_R(\hat{a}_x^\dagger\,\hat{a}_x+\hat{a}_y^\dagger\,\hat{a}_y+1)
+\frac{\omega_L}{2}\,\hat{L}_z\,,
\label{eq1a}
\end{equation}
with $\hat{L}_z=i\hbar(\hat{a}_y^\dagger\,\hat{a}_x-\hat{a}_x^\dagger\,\hat{a}_y)$ being the 
only operator that is non-diagonal in the annihilation and creation operators in the 
above expression. Introducing the 'circular polarisation' operators 
$\hat{a}_+,\hat{a}_-, \hat{a}_+^\dagger,
\hat{a}_-^\dagger$ through $\hat{a}_x=\frac{1}{\sqrt{2}}\,(\hat{a}_++\hat{a}_-)$,
$\hat{a}_y=\frac{i}{\sqrt{2}}\,(\hat{a}_+-\hat{a}_-)$, 
$\hat{a}_x^\dagger=\frac{1}{\sqrt{2}}\,(\hat{a}_+^\dagger+\hat{a}_-^\dagger)$,
$\hat{a}_y^\dagger=-\frac{i}{\sqrt{2}}\,(\hat{a}_+^\dagger-\hat{a}_-^\dagger)$, 
one has $\hat{L}_z=\hbar(\hat{a}_+^\dagger\,\hat{a}_+-
\hat{a}_-^\dagger\,\hat{a}_-)$ and one can write $\hat{H}_0$ as
\begin{equation}
\hat{H}_0=\hbar\,\omega_+\hat{a}_+^\dagger\,\hat{a}_++\hbar\,
\omega_-\hat{a}_-^\dagger\,\hat{a}_-\,+\hbar\,\omega_R\,,
\label{eq2a}
\end{equation}
where $\omega_+=\omega_R+\frac{\omega_L}{2}$, $\omega_-=\omega_R-\frac{\omega_L}{2}$. 
The energy levels are now given in terms of
the occupation numbers of the modes $+,-$ by $E=\hbar\,\omega_+n_++\hbar\,\omega_-n_- 
+\hbar\,\omega_R$. 

We will now compute the dielectric permitivity and optical conductivity of the system
in the quantum regime by considering the Hamiltonian $\hat{H}(t)$ of the system 
interacting with an homogeneous time-dependent electric field, as given in 
section \ref{secB}. Expressing the operators $\hat{x}$ and $\hat{y}$ in terms of 
$\hat{a}_+,\hat{a}_-, \hat{a}_+^\dagger,\hat{a}_-^\dagger$, we have
the following expression for the interaction Hamiltonian $\hat{H}_1(t)$,
\begin{eqnarray}
\hat{H}_1(t)&=&e\,
(\hbar/2m\omega_R)^{1/2}\,[\,{\cal E}_{+}(t)\hat{a}_++{\cal E}_-(t)\hat{a}_+^\dagger+
{\cal E}_-(t)\hat{a}_-+{\cal E}_+(t)\hat{a}_-^\dagger\,]\,,
\label{eq3a}
\end{eqnarray}
where ${\cal E}_{+}(t)=\frac{1}{\sqrt{2}}(E_x(t)+iE_y(t))$ and 
${\cal E}_{-}(t)=\frac{1}{\sqrt{2}}(E_x(t)-iE_y(t))$.

If $\ket{\psi_t}$ is a solution of the time-dependent Schr\"odinger
equation, one defines, as above, the state vector $\ket{\phi_t}=e^{\frac{i}{\hbar}\,H_0t}\,
\ket{\psi_t}$, in the interaction representation, such that
the two vectors coincide at $t=0$, when the field is turned on. 
Given that in the interaction representation the annihilation and
creation operators contained in $\hat{H}_1(t)$ evolve in time through multiplication by a phase factor 
$e^{\pm i \omega_{\pm}\,t}$, $\hat{H}_1^{int}(t)$ is given by
\begin{eqnarray}
\hat{H}_1^{int}(t)&=&e\,
(\hbar/2m\omega_R)^{1/2}\,[\, {\cal E}_{+}(t)\,
e^{-i\omega_+t}\,\hat{a}_++{\cal E}_-(t)
e^{i\omega_+t}\,\hat{a}_+^\dagger\nonumber\\
&&\mbox{}+{\cal E}_-(t)\,e^{-i\omega_-t}\,\hat{a}_-
+{\cal E}_+(t)\,e^{i\omega_-t} \,\hat{a}_-^\dagger\,]\,.
\label{eq4a}
\end{eqnarray}

One can now write, as in section \ref{secB}, the formal solution
$\ket{\phi_t}=T\,\exp\left(-\frac{i}{\hbar}\,\int_{0}^{t}\,du\,\hat{H}_1^{int}(u)\right)\,
\ket{\psi_{0}}$. The time-ordered operator above is then written using
identity (\ref{eq3_1}), the second term in this product being a
phase factor that can be discarded when computing expectation values.
 
We can write for $\ket{\psi_t}$ the result
\begin{equation}
\ket{\psi_t}=e^{-\frac{i}{\hbar}\,\hat{H}_0t}\,
e^{(z_+(t)\hat{a}_+^\dagger-z^*_+(t)\hat{a}_+)+(z_-(t)\hat{a}_-^\dagger-z^*_-(t)\hat{a}_-)}\,\ket{\psi_{0}}\,,
\label{eq5a}
\end{equation}
with $z_\pm(t)$ being given by 
\begin{eqnarray}
z_+(t)&=&-\frac{ie}{\sqrt{2m\hbar\omega_R}}\,\int^{t}_{0}\,du\,{\cal
  E}_{-}(u)\,e^{i \omega_{+}\,u}\,,
\label{eq5a.1}
\\
z_-(t)&=&-\frac{ie}{\sqrt{2m\hbar\omega_R}}\,\int^{t}_{0}\,du\,{\cal
  E}_{+}(u)\,e^{i \omega_{-}\,u}\,,
\label{eq5a.2}
\end{eqnarray}
and where we have discarded the phase factor referred above. The operator
$\hat{D}(z_{+},z_{-})=e^{(z_{+}(t)\hat{a}_{+}^\dagger-z^*_{+}(t)\hat{a}_{+})+(z_{-}(t)\hat{a}_{-}^\dagger-z^*_{-}(t)
\hat{a}_{-})}$ is the displacement operator for the annihilation and creation
operators, i.e. $\hat{D}^\dagger\,\hat{a}_{\pm}\, 
\hat{D}=\hat{a}_{\pm}+z_{\pm}$, $\hat{D}^\dagger\,\hat{a}^{\dagger}_{\pm}\,
\hat{D}=\hat{a}_{\pm}^\dagger+z_{\pm}^*$.
Therefore, using the above representation of $\ket{\psi_t}$, one can show 
that the average value of any operator 
$\hat{A}(\hat{a}_{+},\hat{a}^{\dagger}_{+},\hat{a}_{-},
\hat{a}^{\dagger}_{-})$, in the Schr\"odinger representation, is
given by
\begin{eqnarray}
\langle\,\hat{A}\,\rangle_t&=&\bra{\psi_t}\,\hat{A}(\hat{a}_{+},
\hat{a}^{\dagger}_{+},\hat{a}_{-},\hat{a}^{\dagger}_{-})\,\ket{\psi_t}\nonumber\\
&=&\bra{\psi_{0}}\,\hat{D}^\dagger
\,\hat{A}(\hat{a}_{+}\,e^{-i\omega_+t},\hat{a}^{\dagger}_{+}\,
e^{i\omega_+t},\hat{a}_{-}\,e^{-i\omega_-t},
\hat{a}^{\dagger}_{-}\,e^{i\omega_-t})\,\hat{D}\,\ket{\psi_{0}}\nonumber\\
&=&\bra{\psi_{0}}\,\hat{A}[(\hat{a}_{+}+z_+(t))e^{-i\omega_+t},
(\hat{a}^{\dagger}_{+}+z_+^*(t))e^{i\omega_+t},\nonumber\\
&&\mbox{}(\hat{a}_{-}+z_-(t))e^{-i\omega_-t},
(\hat{a}^{\dagger}_{-}+z_-^*(t))e^{i\omega_-t}]\, \ket{\psi_{0}}\,.
\label{eq6a}
\end{eqnarray}
In particular, if $\hat{A}=\beta^*_+\hat{a}_{+}+\beta_+\hat{a}^{\dagger}_{+}+
\beta^*_-\hat{a}_{-}+\beta_-\hat{a}^{\dagger}_{-}$, i.e. for a linear function
of the annihilation or creation operators, such as $\hat{x},\hat{y},\hat{p}_x,\hat{p}_y$,
which is the case that will concerns us below, one has that
\begin{eqnarray}
\delta\langle\,\hat{A}\,\rangle_t&=&
\beta^*_+z_+(t)\,e^{-i\omega_+t}+\beta_+z_+^*(t)\,e^{i\omega_+t}
+\beta^*_-z_-(t)\,e^{-i\omega_-t}+\beta_-z_-^*(t)\,e^{i\omega_-t}\,,
\label{eq7a}
\end{eqnarray}
where $\delta\langle\,\hat{A}\,\rangle_t=
\langle\,\hat{A}\,\rangle_t-\langle\,\hat{A}\,\rangle_t^0$ is the 
difference between the average value in presence and absence of the
applied electric field. 

Using this result, one can easily show that the induced polarisation in the
system $P_x(t)=-e\,\delta\langle\,\hat{x}\,\rangle_t$,
$P_y(t)=-e\,\delta\langle\,\hat{y}\,\rangle_t$, is given by 
$P_i(t)=\int^{t}_{0}\,du\,\chi_{ij}(t-u)\,E_j(u)$, where the permitivity of the system
is given by
\begin{eqnarray}
\chi_{xx}(t)&=&\chi_{yy}(t)=\frac{e^2}{2m\omega_R}(\sin(\omega_+t)+\sin(\omega_-t))\,,
\label{eq7a.1}\\
\chi_{xy}(t)&=&-\chi_{yx}(-t)=\frac{e^2}{2m\omega_R}(\cos(\omega_+t)-\cos(\omega_-t))\,.
\label{eq7a.2}
\end{eqnarray}
The first equality between the permitivities follows from rotational 
invariance around the $z$ axis, the second
from linear response theory.

The induced current $j_x(t)=-e\,\delta\langle\,\hat{v_x}\,\rangle_t$, $j_y(t)=
-e\,\delta\langle\,\hat{v_y}\,\rangle_t$, is related to $P_x(t)$ and $P_y(t)$, by
$j_x(t)=\dot{P}_x(t)$, $j_y(t)=\dot{P}_y(t)$ and given that $\chi_{ij}(0)=0$,
one easily obtains for the conductivity, defined by $j_i(t)=\int^{t}_{0}\,du\,\sigma_{ij}(t-u)\,E_j(u)$,
the result $\sigma_{ij}(t)=\dot{\chi}_{ij}(t)$. Hence, 
\begin{eqnarray}
\sigma_{xx}(t)&=&\sigma_{yy}(t)=\frac{e^2}{2m\omega_R}(\omega_+\cos(\omega_+t)+\omega_-\cos(\omega_-t))\,,
\label{eq7a.3}
\\
\sigma_{xy}(t)&=&\sigma_{yx}(-t)=-\frac{e^2}{2m\omega_R}(\omega_+\sin(\omega_+t)-\omega_-
\sin(\omega_-t))\,.
\label{eq7a.4}
\end{eqnarray}
Obviously and as above, one can also compute it from the definition
of $\hat{v}_x=\frac{1}{m}\left(\hat{p}_x-\frac{eB}{2}\hat{y}\right)$, $\hat{v}_y=\frac{1}{m}
\left(\hat{p}_y+\frac{eB}{2}\hat{x}\right)$, since all the operators involved are linear
in the annihilation and creation operators. Note that, since $\omega_++\omega_-=2\omega_R$,
$\sigma_{xx}(0)=\sigma_{yy}(0)=e^2/m$, a result which agrees with the {\it f-sum} rule 
for a single quantum particle. It is interesting to consider this system in two limits, 
namely $B\rightarrow 0$ (simple harmonic oscilator) and $\omega_0\rightarrow 0$ 
(particle in a magnetic field). In the
first case, $\omega_+=\omega_-=\omega_0$ and one obtains $\sigma_{xx}(t)=\sigma_{yy}(t)=
\frac{e^2}{m}\cos(\omega_0t)$, $\sigma_{xy}(t)=\sigma_{yx}(-t)=0$ (this result is obvious, 
given the
lack of transverse response if $B=0$). In the second case, $\omega_+=\omega_L$,
$\omega_-=0$, $\omega_R=\omega_L/2$. One has that $\sigma_{xx}(t)=\sigma_{yy}(t)=
\frac{e^2}{m}\cos(\omega_Lt)$, $\sigma_{xy}(t)=\sigma_{yx}(-t)=-\frac{e^2}{m}\sin(\omega_Lt)$.

Finally, let us consider the case in which a constant electric field is turned on at $t=0$. 
In that case, one obtains at large times $t\rightarrow \infty$ that the current 
$j_i=\tilde{\sigma}_{ij}(s\rightarrow 0^+)\,E_j$, where $\tilde{\sigma}_{ij}(s)$ 
is the Laplace transform of $\sigma_{ij}(t)$. Performing the integrals, one obtains 
$\tilde{\sigma}_{xx}(0)=\tilde{\sigma}_{yy}(0)=0$.
The limit $s\rightarrow 0$ requires a bit of care in the transverse conductivity case, 
since $\tilde{\sigma}_{xy}(s)=-\tilde{\sigma}_{yx}(s)=-\frac{e^2}{2m\omega_R}
\left(\frac{\omega_+^2}{\omega_+^2+s^2}-
\frac{\omega_-^2}{\omega_-^2+s^2}\right)$. We obtain $\tilde{\sigma}_{xy}(0)
=-\tilde{\sigma}_{yx}(0)=0$ if $\omega_0\neq 0$. However, we obtain
$\tilde{\sigma}_{xy}(0)=-\tilde{\sigma}_{yx}(0)=-e/B$ in the $\omega_0=0$ 
case ($\omega_-=0$). This result is physically simple to understand if one realises 
that, when a harmonic force is present, a constant electric field merely displaces 
the force center, whether a constant magnetic field is present or
not (a shift in the origin of the coordinates merely contributes a constant term to 
the vector potential, that can be simply gauged away). Therefore, one
will not observe a response of the velocity to the electric field in
that case.  However, in the absence of an harmonic force, the electric field 'pulls' on the
giration radius center as if it were a free particle and one does observe a transverse
response.

One should again note that the quantum and classical results obtained for the susceptibilities
computed above and those obtained from the classical equations
of motion (\ref{eq7}-\ref{eq10}) are identical and, moreover, that the response
to the electric field is purely linear. This result follows from the fact
that the classical equations of motion and their quantum counterparts, the 
Ehrenfest equations, are linear and can therefore be solved with
respect to the field and the initial conditions. In this respect, the system behaviour is
trivial. However, the equality of results between the classical and quantum cases is limited to
operators that are linear combinations of the coordinates and momenta. In the case
of non-linear operators, one can still use the methods discussed in this Appendix to study
their time evolution. Furthermore, if one is interested in the evolution of wave-functions,
as discussed in the main text, one should keep the phase factors that
were discarded in the computation of average values. 

\section{The Infeld-Pleba\'{n}ski identity}
\label{appendixB}
We give here an elementary demonstration of the relation of 
Infeld and Pleba\'nski \cite{Infeld55}. If $\hat{S}_r
=e^{\frac{i}{2\hbar}\,r\,(\hat{\vecb{p}}\cdot\hat{\vecb{r}}+
\hat{\vecb{r}}\cdot\hat{\vecb{p}})}$ is the squeezing
operator introduced above, it is easy to show \cite{Stoler71} that
$\hat{S}_r\,\hat{\vecg{r}}\,\hat{S}_r^\dagger=\hat{\vecg{r}}\,e^r$,
$\hat{S}_r\,\hat{\vecg{p}}\,\hat{S}_r^\dagger=\hat{\vecg{p}}\,e^{-r}$, i.e.
$\hat{S}_r$ is a scale transformation operator that preserves 
the volume of phase space. Using these identities, one can show that 
\begin{eqnarray}
\hat{S}_r\,\ket{\vecg{r}}&=&e^{-rd/2}\,\ket{\vecg{r}\,e^{-r}}\,,
\label{eqB1}\\
\hat{S}_r\,\ket{\vecg{p}}&=&e^{rd/2}\,\ket{\vecg{p}\,e^{r}}\,,
\label{eqB2}
\end{eqnarray}
where $d$ is the space dimension (two in this case).

We now wish to consider the wave-function of a squeezed state evolving
under the isotropic harmonic oscillator $\hat{h}_0$, i.e. the matrix
element $\bra{x,y}\,\,e^{-\frac{i}{\hbar}\,\hat{h}_0t}\,\hat{S}_r\,\ket{\psi_0}$ (the discussion
in momentum space is completely analogous). Inserting a complete
set of position eigenstates, one can write this quantity as
\begin{equation}
\overline{\psi}(x,y,t)=\int_{-\infty}^{+\infty}\int_{-\infty}^{+\infty}\,dx_0\,dy_0\,K_r(x,y,x_0,y_0;t)\,\psi_0(x_0,y_0)\,,
\label{eqB3}
\end{equation}
where $K_r(x,y,x_0,y_0;t)=\bra{x,y}\,\,e^{-\frac{i}{\hbar}\,\hat{h}_0t}\,\hat{S}_r\,\ket{x_0,y_0}$. Using
the identity (\ref{eqB1}), one has that
\begin{equation}
K_r(x,y,x_0,y_0;t)=e^{-r}\,K(x,y,x_0\,e^{-r},y_0\,e^{-r};t)\,,
\label{eqB3_1}
\end{equation}
where $K(x,y,x_0,y_0;t)=\bra{x,y}\,\,e^{-\frac{i}{\hbar}\,\hat{h}_0t}\,\ket{x_0,y_0}$ 
is the isotropic harmonic oscillator propagator. Note, however, that (\ref{eqB3_1}) is valid 
for an arbitrary one-particle Hamiltonian. The harmonic oscillator propagator is given 
by \cite{Feynman65}
\begin{equation}
K(x,y,x_0,y_0;t)=\frac{m\,\omega_R}{2\pi i\hbar\sin(\omega_Rt)}
\exp\left(\frac{im\omega_R}{2\hbar}\left(\,\cot(\omega_Rt)(x^2+y^2+x_0^2+y_0^2)
-\frac{2(\,x\,x_0+y\,y_0\,)}{\sin(\omega_Rt)}\,\right)\right)\,.
\label{eqB4}
\end{equation}
Now, introducing the scaled variables $\tilde{x}=x/\mu_r(t)$,
$\tilde{y}=y/\mu_r(t)$, with $\mu_r(t)=\sqrt{e^{-2r}\cos^2(\omega_Rt)+e^{2r}\sin^2(\omega_Rt)}$, and
$\tau_r=\frac{1}{\omega_R}\,\arctan(e^{2r}\tan(\omega_Rt))$, one can
show, using equation (\ref{eqB3_1}), that 
\begin{eqnarray}
K_r(x,y,x_0,y_0;t)\!\!&=&\!\!\mu_r^{-1}(t)e^{\frac{im\omega_R\mu_r^{-2}(t)
\sinh(2r)\sin(2\omega_Rt)}{2\hbar}\,(x^2+y^2)}
K\left(\frac{x}{\mu_r(t)},\frac{y}{\mu_r(t)},x_0,y_0;\tau_r\right).
\label{eqB5}
\end{eqnarray}
Substituting (\ref{eqB5}) in (\ref{eqB3}), one obtains 
\begin{equation}
\overline{\psi}(x,y,t)=\mu_r^{-1}(t)\,e^{\frac{im\omega_R\mu_r^{-2}(t)}{2\hbar}
\,\sinh(2r)\sin(2\omega_Rt)\,(x^2+y^2)}\,
\psi\left(\frac{x}{\mu_r(t)},\frac{y}{\mu_r(t)},\tau_r\right)\,,
\label{eqB6}
\end{equation}
which is the Infeld-Pleba\'nski relation used in the main text. Since
the expression for the propagator in momentum space is entirely
analogous to (\ref{eqB4}), the steps are identical to those above,
except that $r$ is replaced by $-r$. One obtains
\begin{equation}
\overline{\psi}(p_x,p_y,t)=\mu_{-r}^{-1}(t)\,e^{-\frac{i\mu_{-r}^{-2}(t)}{2\hbar
    m\omega_R}
\,\sinh(2r)\sin(2\omega_Rt)\,(p_x^2+p_y^2)}\,
\psi\left(\frac{p_x}{\mu_{-r}(t)},\frac{p_y}{\mu_{-r}(t)},\tau_{-r}\right)\,.
\label{eqB7} 
\end{equation}

\begin{figure}[htbp]
\begin{center}
\includegraphics*[width=8cm]{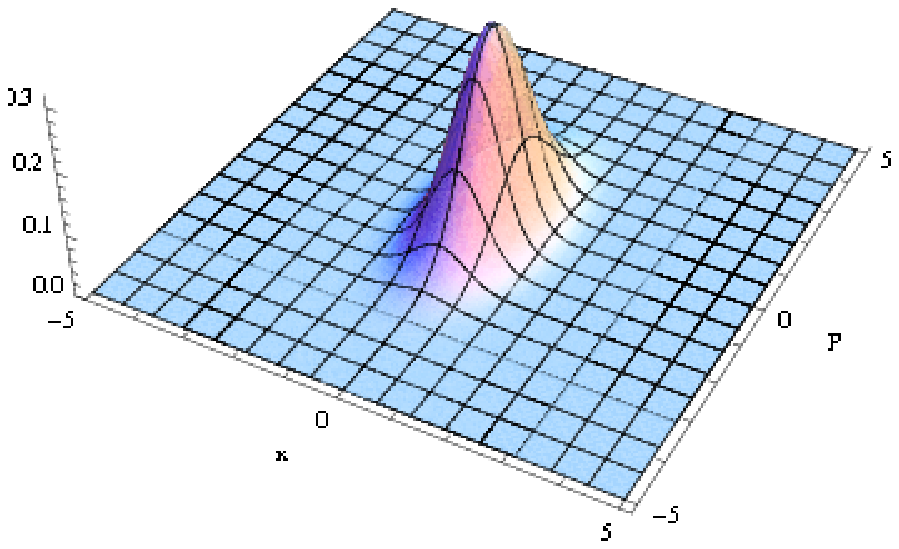}
\end{center}
\vspace{3mm}
\caption{\label{F1}}
\end{figure}

\newpage
\noindent{{\bf  Figure Captions}\\
{\bf Figure 1.}
$x,p_x$ section of Wigner function of squeezed-coherent state,
as given by equation (\ref{eq17}) at $t=0$, evolving
under the Fock-Darwin Hamiltonian. The length of the $x$, $p_x$ axis 
is, respectively, $10^{-8}$ m and $10^{-26}$ Kg m s$^{-1}$
and the Wigner function was multiplied by
$\hbar^2$.\\
{\bf Figure 2.}\\
Same as \ref{F1} at time $t=\pi/(2\omega_R)$.
\\
{\bf Figure 3.}\\
Same as \ref{F1} at time $t=3\pi/(4\omega_R)$.
\\
{\bf Figure 4.}\\
Same as \ref{F1} at time $t=5\pi/(4\omega_R)$.}
\newpage
\newpage
\begin{figure}[htbp]
\begin{center}
\includegraphics*[width=8cm]{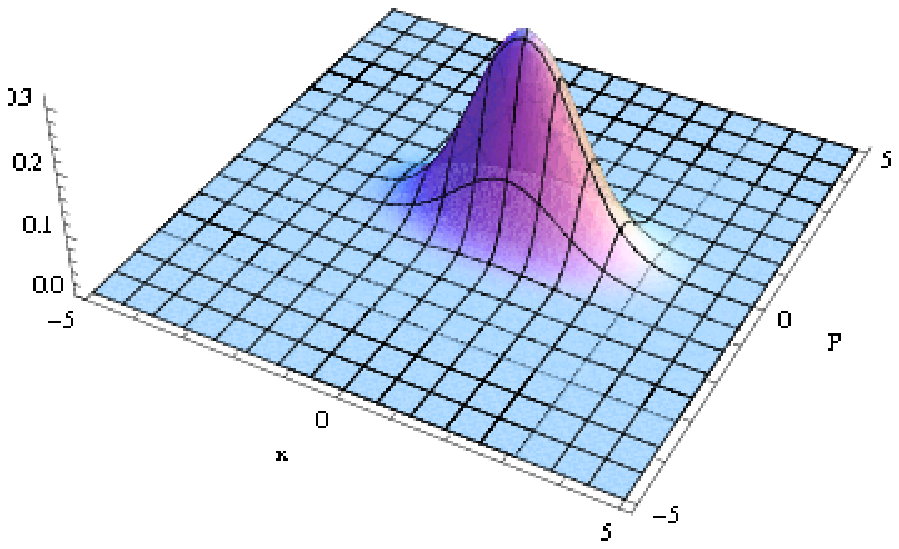}
\end{center}
\vspace{3mm}
\caption{\label{F2}}
\end{figure}
\newpage
\begin{figure}[htbp]
\begin{center}
\includegraphics*[width=8cm]{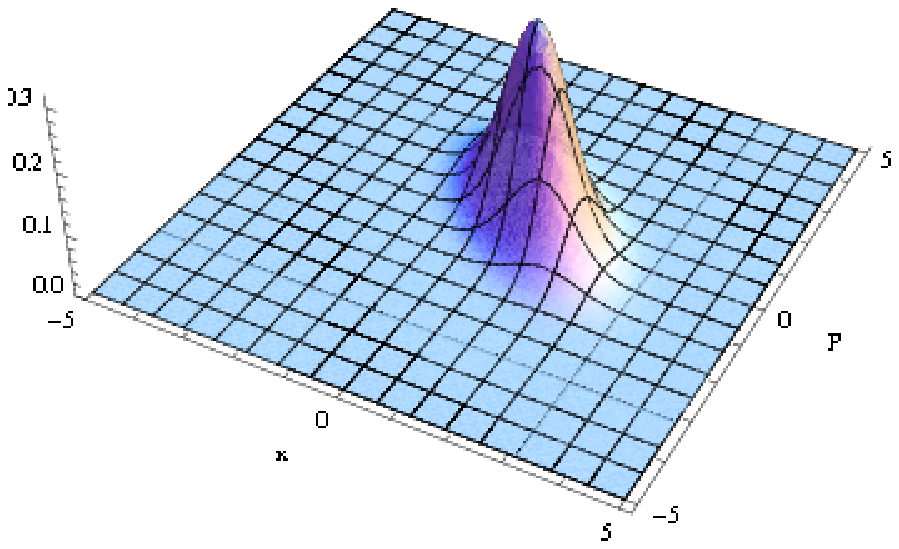}
\end{center}
\vspace{3mm}
\caption{\label{F3}}
\end{figure}
\newpage
\begin{figure}[htbp]
\begin{center}
\includegraphics*[width=8cm]{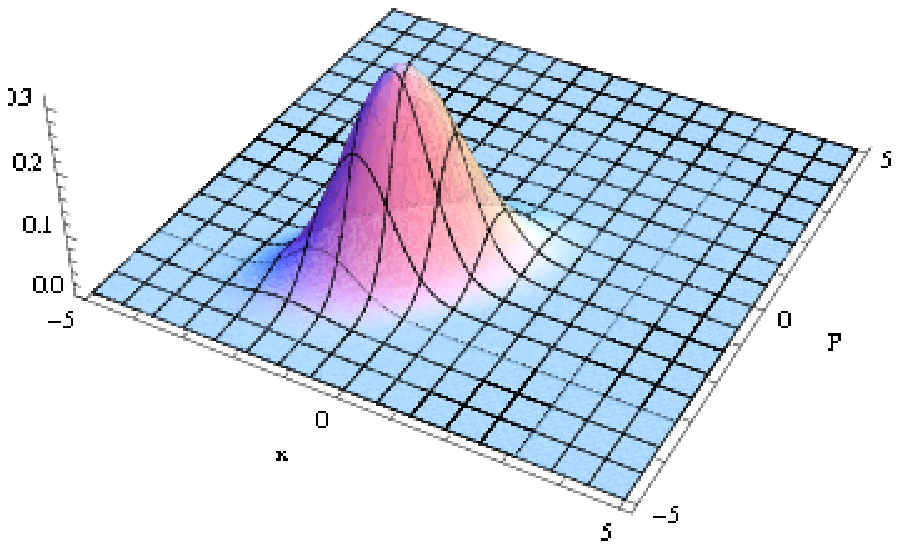}
\end{center}
\vspace{3mm}
\caption{\label{F4}}
\end{figure}
\end{document}